\documentclass[aps,showpacs,twocolumn,superscriptaddress]{revtex4}

\usepackage{amsmath}
\usepackage{amssymb}
\usepackage{amscd}
\usepackage{bm}
\usepackage{wasysym}
\usepackage[ansinew]{inputenc}
\usepackage[T1]{fontenc}
\usepackage{ae,aecompl}
\usepackage{hyperref}

\usepackage[rightcaption]{sidecap}

\usepackage[dvips]{graphicx}
\DeclareGraphicsExtensions{.eps}

\newcommand{\be}{\begin{equation}}
\newcommand{\ee}{\end{equation}}
\newcommand{\bea}{\begin{eqnarray}}
\newcommand{\eea}{\end{eqnarray}}

\usepackage{color}
\definecolor{red}{rgb}{1,0,0}

\begin{document}
\bibliographystyle{apsrev}

\title{Mean-field dynamics of a Bose-Hubbard chain coupled to a non-Markovian environment}

\author{G. Kordas}
\affiliation{National and Kapodistrian University of Athens, Physics Department, Nuclear \& Particle Physics Section Panepistimiopolis, Ilissia 15771 Athens, Greece}

\author{G. E. Pavlou}
\affiliation{National and Kapodistrian University of Athens, Physics Department, Nuclear \& Particle Physics Section Panepistimiopolis, Ilissia 15771 Athens, Greece}

\author{A.I. Karanikas}
\affiliation{National and Kapodistrian University of Athens, Physics Department, Nuclear \& Particle Physics Section Panepistimiopolis, Ilissia 15771 Athens, Greece}

\date{\today }

\begin{abstract}
We study the dynamics of an interacting Bose-Hubbard chain coupled to a non-Markovian environment. Our basic tool is  the reduced generating functional expressed as a path integral over spin coherent states.  We calculate the leading contribution to the corresponding effective action, and by minimizing it, we derive mean-field equations that can be numerically solved. With this tool at hand, we examine the influence of the system's initial conditions and interparticle interactions on the dissipative dynamics. Moreover, we investigate the presence of memory effects due to the non-Markovian environment.
\end{abstract}

\maketitle

\section{Introduction}
\label{sec:intro}
Open quantum many-body systems have gained significant attention due to their importance in quantum information processing, experiments with ultracold atoms and other areas of scientific and technological interest. An environment is always present and it usually has destructive influence on quantum coherence. However, there are plenty of works that prove that a carefully designed environment can be used to control the many-body dynamics~\cite{Kor15}. It is thus of great importance to understand all the aspects of such systems.

The usual treatment of open quantum many-body systems is based on a Lindblad master equation~\cite{Kor15,Breuer2007}, which assumes a memoryless environment. However, if the environment is structured or if the coupling between system and environment is strong, then this approach is not valid. For quadratic Hamiltonians it is possible to use non-Markovian master equations, which can be derived with the help of the Feynman-Vernon formalism~\cite{an07,tu08,Chen13,Shi13,an2009}. In addition to that non-Markovian approaches such as quantum jumps or quantum trajectories~\cite{jing11,jing13,piilo08,breuer04} can also be used. The case of non-quadratic Hamiltonians, such as the interacting Bose-Hubbard (BH) \cite{Hubbard1963}, are by far more complicated. The difficulties begin already when one tries to express the problem in the context of the Feynman-Vernon influence functional. As it has been shown~\cite{wilson11} there are inconsistencies in the definition of the coherent state path integrals for such Hamiltonians, making the derivation of the influence functional problematic. Only recently, a simple recipe to define bosonic~\cite{Kor14} and spin~\cite{Kor16} coherent-state path integrals has been given. However, even if the influence functional is well defined, it is impossible to derive a non-Markovian master equation due to the presence of interactions.

In this work we make a first step to understanding the dynamics of a non-quadratic many-body Hamiltonian coupled to a non-Markovian environment. In Sec.~\ref{sec:2} we present the model we are going to use: a BH chain coupled to a non-Markovian bath of harmonic oscillators. After that we write the thermal generating functional in the language of coherent-state path integrals and we integrate out the degrees of freedom of the environment to obtain the reduced generating functional. Since we are interested on the influence of vacuum fluctuations to the evolution we examine the zero-temperature limit. In this way we study a dissipative mechanism which is important in optical systems, where the modes are inevitably coupled to the vacuum of the electromagnetic field~\cite{an07}. In Sec.~\ref{sec:3}, we use the effective action approach to derive mean-field equations of motion, in the form of a non-Markovian discrete nonlinear Schr\"{o}dinger (NMDNLS) equation. 

In Secs.~\ref{sec:4} and~\ref{sec:5}, we study the dynamics of two small but interesting systems, the two-site and the four-site BH models, respectively. In the two-site model, we compare the dynamics between different types of non-Markovian environments and their respective Markovian limit. For different interaction strengths we find that the particle losses are larger for larger interactions between the particles and we explain the phenomenon. Moreover, we investigate the initial conditions leading to maximum or minimum losses at a given instant. We find that if we begin from the unstable fixed point of the dissipative DNLS, there are no losses and we discuss the phenomenon. Extending our discussion to the four-site system we find that the essential property of the zero-loss initial conditions  is their underlying $Z_2$ symmetry. Finally, we study a purely non-Markovian effect: the return of particles back to the system from the environment due to memory effects. 

\section{The reduced generating functional}
\label{sec:2}
In this work we will study the influence of a non-Markovian environment on the dynamics of a BH chain. The BH Hamiltonian has been used successfully to describe the dynamics of a great variety of systems, from ultracold atoms in optical lattices~\cite{morsch06} to discrete optical systems~\cite{longhi11}. The BH Hamiltonian, that describes our system, is given by
\begin{eqnarray}\label{eq:BH}
\nonumber \hat{H}_{\rm S} &=& \sum_{j=1}^M \varepsilon_j \hat{\alpha}^\dag_j \hat{\alpha}_j + \frac{U}{2}\sum_{j=1}^M\hat{\alpha}^\dag_j \hat{\alpha}^\dag_j \hat{\alpha}_j\hat{\alpha}_j \\
&& - J\sum_{j=1}^{M-1}(\hat{\alpha}^\dag_{j} \hat{\alpha}_{j+1} + \hat{\alpha}^\dag_{j+1} \hat{\alpha}_{j}),
\end{eqnarray}
where $\hat{\alpha}_j$, $\hat{\alpha}_j^\dag$ are bosonic annihilation and creation operators, $\varepsilon_j$ are the on-site energies, $J$ is the tunneling strength, $U$ is the interparticle interaction strength, while we set $\hbar=1$.

We shall adopt the usual approach and we are going to simulate the environment as an infinite collection of harmonic oscillators~\cite{cald83,raja15,Weiss2012}
\begin{equation}\label{eq:R}
\hat{H}_R = \sum_k E_k \hat{R}_k^\dag \hat{R}_k
\end{equation}
where $\hat{R}_k$ and $\hat{R}_k^\dag$ are annihilation and creation bosonic operators for the $k$-th oscillator.

Finally, we consider the interaction between the system and the environment to be linear
\begin{equation}\label{eq:int}
\hat{H}_{\rm I} = \sum_{j,k}( \gamma_{j,k} \hat{R}_k \hat{\alpha}_j^\dag + \gamma_{j,k}^* \hat{R}_k^\dag \hat{\alpha}_j).
\end{equation}

Thus, the Hamiltonian that describes the total system has the form
\begin{equation}\label{eq:Htot}
\hat{H} = \hat{H}_{\rm S} + \hat{H}_{\rm R} + \hat{H}_{\rm I}.
\end{equation}

At this point we shall introduce our basic mathematical tool: the functional that, at a finite temperature $\beta=1/kT$, generates correlation functions pertaining to  the composite system (\ref{eq:Htot}). We shall express this functional as a path integral over the space spanned by the coherent basis. We introduce the notation $|\bm{a}\rangle = |a_{1},...,a_{M}\rangle$ and $|\bm{r}\rangle = |r_{1},...,r_{k},...\rangle$ for the over-completed bases pertaining to the system and the environment respectively. For the composite system we use the notation $|\bm{z}\rangle = |\bm{a},\bm{r}\rangle$, while the completeness relation can be casted in the abbreviated form
\begin{eqnarray}
\nonumber\int d^2 \bm{z}|\bm{z}\rangle\langle\bm{z}| &\equiv& \prod_{j\in{\rm S}}\int \frac{da_{j}da_j^*}{2\pi i} |a_j\rangle\langle a_j| \times \\
&& \times \prod_{k\in{\rm R}}\int \frac{dr_{k}dr_{k}^*}{2\pi i} |r_k\rangle\langle r_k| = \hat{I}.\label{eq:pi6}
\end{eqnarray}
Path integration in the complexified phase space is ultimately connected with the underlying time-slice structure~\cite{Kleinert2009,kochetov12}. The continuum limit has to take properly into account this structure in order to avoid inconsistencies~\cite{wilson11}.
One way to avoid such problems is based on the introduction of the proper  ``classical'' Hamiltonian $H^F$ that weighs paths in the space spanned by the coherent states. This classical Hamiltonian can be obtained from the quantum one via a simple route~\cite{Kor14}
\begin{equation}\label{eq:pi11}
\hat{H}(\hat{\alpha}^\dag,\hat{\alpha})\rightarrow \hat{H}(\hat{p},\hat{q})\rightarrow H^F(p,q) \rightarrow H^F(a^*,a).
\end{equation}
The first step in this chain is the replacement of the creation and annihilation operators by
the corresponding quadratures  (``momentum" and ``position" operators). Next, one passes to the classical Hamiltonian appearing in the Feynman phase space integral and eventually performs a canonical change of variables: $q=(a^*+a)/\sqrt{2}$ and $q=i(a^*-a)/\sqrt{2}$.

In this way the generating functional for the composite system at a finite temperature is defined as follows:
\begin{eqnarray}
\nonumber Z[\mathcal{J}] &=& \frac{1}{Z(\beta)} \int d^2 \bm{w} \mathop{\int \mathcal{D}^2 \bm{z}}\limits_{\substack{\bm{z}^*(\beta)=\bm{w}^* \\ \bm{z}(0)=\bm{w}}} \exp \left\{ -\Gamma(\bm{z}^*,\bm{z})   \right. \\
\nonumber && -\int_0^\beta d\tau \left[\frac{1}{2} (\bm{z}^*\cdot\dot{\bm{z}}- \bm{z}\cdot\dot{\bm{z}}^*)+H^F(\bm{z}^*,\bm{z})\right]\\
&& \left.- \int_0^\beta d\tau (\bm{a}^* \cdot\bm{\mathcal{J}} + \bm{\mathcal{J}}^*\cdot\bm{a})\right\}.\label{eq:pi12}
\end{eqnarray}
Here $\bm{\mathcal{J}}$ is an auxiliary source term, while the integration over $\bm{w}=(\bm{a}^{(0)};\bm{r}^{(0)})$ takes care of the periodic boundary conditions. The classical Hamiltonian, $H^F=H^F_S+H^F_R+H^F_I$, has been constructed through the rule (\ref{eq:pi11}) and its terms read as follows:
\begin{eqnarray}\label{eq:pi16}
\nonumber H^F_{\rm S} &=& \sum_{j=1}^M(\varepsilon_j + U)|a_{j}|^2 - J\sum_{j=1}^{M-1}(a_{j}^*a_{j+1} + a_{j+1}^*a_{j}) \\
&& + \frac{U}{2} |a_{j}|^4 + \sum_{j=1}^M \left(\varepsilon_j + \frac{3U}{8} \right),
\end{eqnarray}
\begin{equation}
H^F_I=\sum_{j,k}(\gamma_{kj}r_ka^*_j + \gamma^*_{kj}r^*_k a_j)
\end{equation}
and
\begin{equation}
H^F_{\rm R} = \sum_k E_k |r_{k}|^2 - \frac{1}{2}\sum_k E_k.
\end{equation}
The boundary factor $\Gamma$ appearing in (\ref{eq:pi12}) reads as follows:
\begin{equation}
\Gamma = |\bm{w}|^2 - \frac{1}{2}(\bm{w}^*\cdot\bm{z}(\beta) + \bm{w}\cdot\bm{z}^*(0)).
\end{equation}
Finally, $Z(\beta)={\rm Tr}(e^{-\beta\hat{H}})=Z[\mathcal{J}=0]$ is the partition function of the total system.

Since we are interested only in quantities pertaining the system $S$, we are going to integrate out the degrees of freedom of the environment, $\bm{r}$, to obtain the reduced generating functional. Due to the fact that it is just a collection of harmonic oscillators and the interaction with the system is linear, the integration can be easily performed. One needs only a change of variables in order to get rid of the boundary conditions: $\bm{r} = \bm{r}^{cl.} + \bm{\eta}_{\rm R}$, $\bm{r}^* = \bm{r}^{cl.*} + \bm{\eta}_{\rm R}^*$ with $\bm{r}^{cl.*}(\beta) = \bm{r}^{(0)*}$ and $\bm{r}^{cl.}(0) = \bm{r}^{(0)}$. The functions
\begin{eqnarray}
 r_{k}^{cl.}(\tau) &=& e^{-E_k\tau}r_{k}^{(0)} \label{eq:pi17a}\\
\nonumber && - \sum_j\gamma_{kj}^* \int_0^\tau d\tau' e^{-E_k(\tau-\tau')}a_{j}(\tau'),\\
r_{k}^{cl.*}(\tau) &=& e^{-E_k(t-\tau)}r_{k}^{(0)*} \label{eq:pi17b}\\
\nonumber && - \sum_j\gamma_{kj} \int_\tau^t d\tau' e^{-E_k(\tau-\tau')}a_{j}^*(\tau')
\end{eqnarray}
have been chosen to enforce stationarity, with respect to the environmental degrees, of the exponent in Eq. (\ref{eq:pi12}). The rest of the calculation is just a quadratic fluctuation integral that can be evaluated by standard means. Its contribution yields an exponential factor $e^{-\tau\sum_k E_k/2}$~\cite{Kor14} that it is exactly canceled by the constant term appearing in $H^F_{\rm R}$. Thus the integration of the environment is encapsulated in the classical solutions Eqs. (\ref{eq:pi17a}) and (\ref{eq:pi17b}), and through them (taking into account that $S_{\rm R}^{cl.}=0$), in the $\Gamma_{\rm R}$ factor. After these explanations one can easily confirm that the reduced generating functional (\ref{eq:pi12}) assumes the form
\begin{eqnarray}\label{eq:18}
 Z[\mathcal{J}] &=& \frac{1}{Z_S(\beta)} \int d^2 \bm{a}^{(0)} \times \\
\nonumber &\times & \mathop{\int \mathcal{D}^2 \bm{a}}\limits_{\substack{\bm{a}^*(\beta)=\bm{a}^{(0)*} \\ \bm{a}(0)=\bm{a}^{(0)} }} e^{ -\Gamma_{\rm S}(\bm{a}^*,\bm{a}) -\int_0^\beta d\tau (\bm{a}^*\cdot\bm{\mathcal{J}} + \bm{\mathcal{J}}^*\cdot\bm{a}) - \tilde{S}[\bm{a}^*,\bm{a}]},
\end{eqnarray}
where
\begin{eqnarray}\label{eq:pi19}
\tilde{S} &=& \int_0^\beta d\tau \left[\frac{1}{2}(\bm{a}^*\cdot\dot{\bm{a}} - \dot{\bm{a}}^*\cdot\bm{a}) + H_S^F\right]\\
\nonumber && - \sum_{j,\ell\in \rm S} \int_0^\beta d\tau \int_0^\tau d\tau' a_{j}^*(\tau) \mu_{j\ell}(\tau-\tau') a_{\ell}(\tau'),
\end{eqnarray}
with $\mu_{j\ell}$ the dissipation kernel:
\begin{equation}
\mu_{j\ell}(\tau-\tau') = \sum_{k\in R}\gamma_{kj}^* \gamma_{k\ell} e^{-E_k(\tau-\tau')}.
\label{eq:mu}
\end{equation}

The functional (\ref{eq:18}) can be used for the generation of thermal correlation functions pertaining to the subsystem. However, in this work, we will use this expression as a mathematical tool to study the zero-temperature, ground-state properties of the subsystem. To this end we define the thermal expectation value of a system's operator $\hat{O}_S$:
\begin{eqnarray}
\nonumber\langle \hat{O}_S(t)\rangle_\beta &\equiv& {\rm Tr}\left\{ \frac{e^{-\beta \hat{H}}}{Z} \hat{O}_S(t)\otimes \hat{I}_R \right\}\\
&=&{\rm Tr}_R\left\{\hat{\rho}_S(\beta)\hat{O}_S(t) \right\},\label{eq:28}
\end{eqnarray}
where
\begin{equation}
\hat{\rho}_S(\beta)\equiv {\rm Tr}_R\left\{ \frac{e^{-\beta \hat{H}}}{Z} \right\}
\end{equation}
is the reduced density matrix of the system. If the system's ground state is unique, the zero temperature limit, $\beta  \to \infty$ , projects Eq. (\ref{eq:28}) on its vacuum expectation value:
\begin{eqnarray}
\nonumber \langle\hat{O}_S(t)\rangle_\beta &\equiv& \sum_{n\in (S+R)}\langle n|\left( \frac{e^{-\beta\hat{H}}}{Z} \hat{O}_S\right)|n\rangle =\\
\nonumber &\mathop{\longrightarrow}\limits_{\beta\rightarrow\infty}& \langle G| \hat{O}_S(t)|G\rangle =\\
&=& {\rm Tr}_S \{\hat{\rho}_S \hat{O}_S(t)\} = \langle \hat{O}_S(t)\rangle,\label{eq:29}
\end{eqnarray}
where $|G\rangle$ is the ground state of the composite system and $\hat{\rho}_S={\rm Tr}_R\{ |G\rangle\langle G| \}$. In case of degeneracy, the zero temperature limit produces an equal probable mixture of all the possible ground states. 

The thermal vacuum expectation values can be derived from (\ref{eq:18}) by functional differentiation and by making the Wick rotation $\tau=it$. For example
\begin{eqnarray}
\nonumber \left.-\frac{\delta \ln Z}{\delta\mathcal{J}_k^*(\tau)}\right|_{\bm{\mathcal{J}}=\bm{0}} &=& \frac{1}{Z_S[0]} \int d^2 \bm{a}^{(0)} \times \\
\nonumber &&\times \mathop{\int \mathcal{D}^2 \bm{a}}\limits_{\substack{\bm{a}^*(\beta)=\bm{a}^{(0)*} \\ \bm{a}(0)=\bm{a}^{(0)} }} e^{-\Gamma(\bm{a}^*,\bm{a}) - \tilde{S}[\bm{a}^*,\bm{a}]} a_k\\
&=& {\rm Tr}_S \{ \hat{\rho}_S(\beta)\hat{a}_k(\tau)\} \equiv \langle \hat{a}_k(\tau) \rangle_\beta.\label{eq:30}
\end{eqnarray}
This result coincides with (\ref{eq:28}) if we make the change $\tau=it$ while the limit $\beta\rightarrow\infty$ produces the result (\ref{eq:29}).

\section{The effective Action and the Mean-Field Approximation}
\label{sec:3}
In this section we shall use the effective action approach to derive equations of motion for  the system's field vacuum expectation values. To begin with we the mean-values
\begin{eqnarray}\label{eq:31}
\langle a_k(\tau)\rangle_\mathcal{J} &=& -\frac{\delta \ln Z}{\delta\mathcal{J}_k^*(\tau)},\\
\nonumber \langle a_k^*(\tau)\rangle_\mathcal{J} &=& -\frac{\delta \ln Z}{\delta\mathcal{J}_k(\tau)}\\
&=& \langle a_k(\tau)\rangle_\mathcal{J}^*.\label{eq:32}
\end{eqnarray}
Note that Eq. (\ref{eq:31}) is not (\ref{eq:30}), since it depends on the source fields $\mathcal{J}$. They are equivalent only in the limit $\mathcal{J}=0$.
Equations (\ref{eq:31}) and (\ref{eq:32}) can, in principle, be solved with regard to the source fields $\mathcal{J}$:
\begin{eqnarray}
\bm{\mathcal{J}} &=& \bm{\mathcal{J}}[\langle \bm{a}(\tau)\rangle_\mathcal{J},\langle \bm{a}(\tau)\rangle_\mathcal{J}^*],\\
\bm{\mathcal{J}}^* &=& \bm{\mathcal{J}}^*[\langle \bm{a}(\tau)\rangle_\mathcal{J},\langle \bm{a}(\tau)\rangle_\mathcal{J}^*].
\end{eqnarray}
The effective action is defined as follows (see Ref.~\cite{Kleinert2009})
\begin{eqnarray}
\nonumber && A[\langle \bm{a}(\tau)\rangle_\mathcal{J},\langle \bm{a}(\tau)\rangle_\mathcal{J}^*] \equiv - \ln Z[\mathcal{J}] \\
&& - \int_0^\beta d\tau (\bm{\mathcal{J}}^* \cdot \langle \bm{a}(\tau)\rangle_\mathcal{J} + \bm{\mathcal{J}} \cdot \langle \bm{a}^*(\tau)\rangle_\mathcal{J}).\label{eq:EA}
\end{eqnarray}
One can easily confirm that
\begin{equation}
\frac{\delta A}{\delta \langle a_k(\tau)\rangle_\mathcal{J}^*}=-\mathcal{J}_k,~\frac{\delta A}{\delta \langle a_k(\tau)\rangle_\mathcal{J}}=-\mathcal{J}_k^*.
\end{equation}
Thus, minimization of the effective action yields equations the solution of which produces the vacuum expectation values (\ref{eq:30}).

As is obvious from the preceding discussion, an exact calculation of the effective action is impossible. However, if the number of particles  is large enough, a systematic approximation in powers of $1/\sqrt N$ is possible. The reason for this can be traced back to the relation $\sum\limits_k { < a_k^{\dag}(0){a_k}(0) > }  = {N_0}$ which calls for the introduction of the rescaled variables $a/\sqrt N_0$. In such a case a large factor $N_0$ appears in the definition of the action $\tilde{S}$ permitting a systematic semiclassical \cite{Kleinert2009,Kleinert13} calculation of the effective action. The first step towards this direction begins from the ``classical" equations of motion:

\begin{equation}\label{eq:33}
\frac{\delta \tilde{S}}{\delta a^{cl.*}_k}=-\mathcal{J}_k,~\frac{\delta \tilde{S}}{\delta a^{cl.}_k}=-\mathcal{J}_k^*.
\end{equation}
The boundary conditions of these equations can be deduced from the integral (\ref{eq:18}).

Skipping all the intermediate steps that can be found elsewhere (see \cite{Kleinert2009}) we find that in leading order:
\begin{eqnarray}
A &\approx& \tilde{S}^{cl.} \approx \int_0^\beta d\tau (\bm{a}^{*cl.}\cdot\dot{\bm{a}}^{cl.} + H_S^{F,cl.}) - \\
\nonumber &&-\sum_{j,\ell}\int_0^\beta d\tau \int_0^\tau d\tau' a_j^{cl.*}(\tau)\mu_{j\ell}(\tau-\tau')a_\ell^{cl.}(\tau').
\label{eq:A}
\end{eqnarray}

In this paper we shall not take into account the quantum corrections in the last equation staying in the mean-field approximation of the problem. By minimizing $\tilde{S}^{cl.}$ we get the following equation for the mean value (\ref{eq:30}):
\begin{equation}
\label{eq:36}
\begin{split}
&\partial_\tau \langle a_k (\tau ) \rangle_\beta + \frac{\partial}{\partial \langle a_k(\tau )\rangle_\beta^*} H_S^F = \\& \sum_j \int_0^\tau d\tau' \langle a_j (\tau')\rangle_\beta \mu_{jk}(\tau-\tau').
\end{split}
\end{equation}

By performing the Wick rotation, $\tau=it$, and taking the zero temperature limit we get the following equation pertaining to the vacuum expectation values (\ref{eq:28}):
\begin{equation}\label{eq:37}
i\partial_t \langle a_k(t)\rangle - \frac{\partial}{\partial \langle a_k(t)\rangle^*} H_S^F = -i\sum_j \int_0^t dt' \langle a_j (t')\rangle \mu_{jk}(t-t').
\end{equation}
Equations (\ref{eq:36}) and (\ref{eq:37}) although formally similar, have different physical interpretation. Equation (\ref{eq:36}) is a diffusion equation and the variable $\tau\in[0,\beta]$ parametrizes local variations of the temperature, while (\ref{eq:37}) is an evolution equation where $t\in[0,\infty)$ is the time. The boundary conditions of the two equations are also different. In (\ref{eq:36}) the boundary conditions are periodic, as are the conditions under which the conditions  the ``classical" Eqs. (\ref{eq:33}) were solved. The boundary conditions accompanying  Eq. (\ref{eq:37}) need not be the same as the corresponding  physical problem is quite different. In the mean field approximation they can be determined through the requirement $\sum\limits_k { < a_k^{\dag}(0){a_k}(0) > }  = {N_0}$.

In explicit form the equations of motion (\ref{eq:37}) read:
\begin{eqnarray}\label{eq:NMDNLS}
 && i\frac{d}{dt} \langle a_k(t)\rangle - (\varepsilon + U)\langle a_k(t)\rangle + J(\langle a_{k+1}(t)\rangle \\
\nonumber &&+ \langle a_{k-1}(t)\rangle)  - U|\langle a_k(t)\rangle|^2 \langle a_k(t)\rangle \\
\nonumber &&= - i \int_0^t dt' \sum_{\ell\in\rm S} \mu_{k\ell}(t-t') \langle a_\ell(t')\rangle.
\end{eqnarray}
The left hand side (lhs) of Eq. (\ref{eq:NMDNLS})  is the well-known discrete non-linear Schr{\"o}dinger equation (DNLS), while the right hand side (rhs) contains a dissipation term which is non-local in time and incorporates all the effects of the non-Markovian environment. Due to the fact that all the lattice sites are inter-connected through the environment, this dissipation term is also non-local in space.

Before proceeding, we shall make the simplifying assumption that the coupling strengths $\gamma_{kj}$ are the same, in magnitude and phase, for all sites: $\gamma_{jk}\equiv\gamma_k$. The structure of the environment is specified by its spectral density:
\begin{equation}
D(E)=\sum_k |\gamma_k|^2\delta(E-E_k)
\end{equation}
At the continuum limit, the usually adopted form ~\cite{an07} for this function reads as follows:
\begin{equation}\label{eq:26}
D(E) = \lambda E \left(\frac{E}{E_{\rm c}}\right)^{n-1} e^{-\frac{E}{E_c}}~~(E > 0).
\end{equation}
Here $\lambda$ is a dimensionless coupling constant and $E_{\rm c}$ is an exponential cutoff. We shall fix the parameters as follows: $\lambda=0.005$ and $E_{\rm c}=30$ \cite{an2009}. Depending on the value of $n$ the environment is classified as sub-Ohmic ($0<n<1$), Ohmic ($n=1$) and super-Ohmic ($n>1$) \cite{an2009}. In what follows we will consider spectral densities only of the form (\ref{eq:26}). The site independent dissipation kernel in the continuum is the Laplace transform of the spectral
density and it can be calculated from Eq.  (\ref{eq:mu}):
\begin{equation}\label{eq:27}
\mu (t  - t') = \int\limits_0^\infty  {dED(E){e^{ - iE(t  - t')}}.} 
\end{equation}

\begin{figure}[tb]
\centering
\includegraphics[width=8.5cm, angle=0]{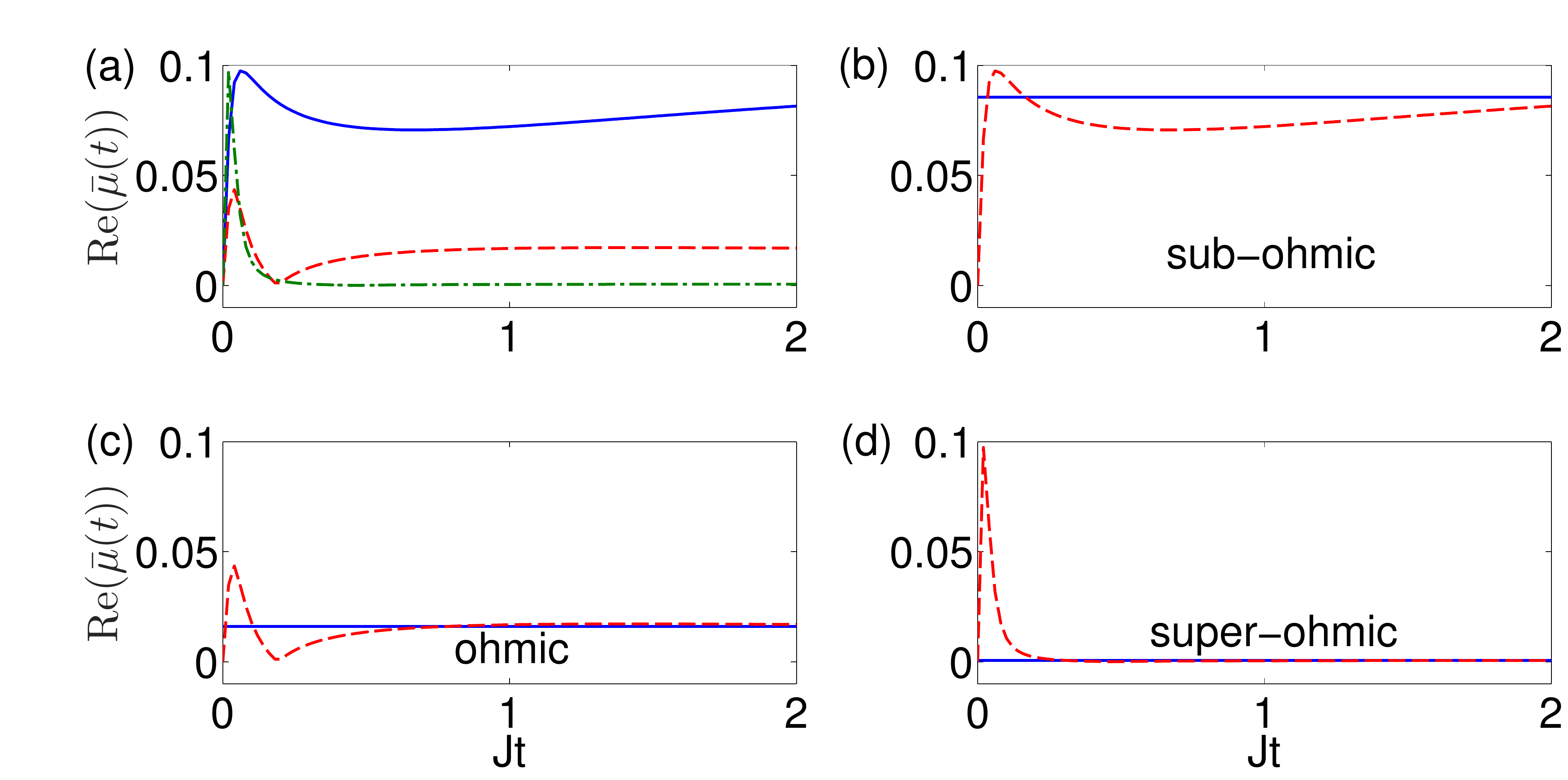}
\caption{\label{fig:0}
(Color online)
Dependence of the real part of the averaged dissipation kernel $\bar\mu (t) $. (a) Comparison between sub-Ohmic, $n=0.5$ (solid blue line), Ohmic, $n=1$ (red dashed line) and super-Ohmic, $n=2$ (green dashed dotted line) non-Markovian environments. (b-d) Comparison between Markovian (red dashed line) and non-Markovian (blue solid line) dynamics for sub-Ohmic ($\Gamma=0.0855J$) (b), Ohmic ($\Gamma=0.0161J$) (c) and super-Ohmic ($\Gamma=5.6 \times {10^{ - 4}}J$) (d) environments. In all cases $UN_0=6J$.
}
\end{figure}

For comparison reasons we shall also examine the mean-field equation at the Markovian limit. To this end we redefine the system's variables as $\left\langle {{a_j}\left( t \right)} \right\rangle  \to \left\langle {{a_j}\left( t \right)} \right\rangle {e^{ - i(\varepsilon  + U)t}}$ (where $\varepsilon  = \max ({\varepsilon _j}),j \in S$) and in what follows we set ${\varepsilon _j} = \varepsilon$. The Markovian case corresponds to the case in which the system's time scale ($\sim 1/\varepsilon$) is much larger than the corresponding environmental characteristic scale. At this limit, in integrals like the one appearing in Eq (\ref{eq:NMDNLS}), we can write

\begin{equation}
\left\langle {{\alpha _j}(t')} \right\rangle  \cong \left\langle {{\alpha _j}(t)} \right\rangle.
\end{equation}
At the Markovian limit the memory effects due to the presence of the environment are absent. This is a natural expectation when the environment is stochastic and much larger than the system itself.  Now the integral on the rhs of Eq. (\ref{eq:NMDNLS}) can be calculated, leading to the result:

\begin{eqnarray}\label{eq:MDNLS}
 && i\frac{d}{dt} \langle a_k(t)\rangle - \Delta_\varepsilon\langle a_k(t)\rangle + J(\langle a_{k+1}(t)\rangle \\
\nonumber &&+ \langle a_{k-1}(t)\rangle)  - U|\langle a_k(t)\rangle|^2 \langle a_k(t)\rangle \\
\nonumber &&= - i \Gamma \sum\limits_{\ell  \in S} {\langle {a_\ell }(t')\rangle }.
\end{eqnarray}

where
\begin{eqnarray}
\Delta_\varepsilon &\equiv& 2(\varepsilon + U) + {\rm Pr.} \int_0^\infty dE \frac{D(E)}{E-(\varepsilon +U)},\\
\Gamma &\equiv&\pi D(\varepsilon + U).
\end{eqnarray}
It is worth noting that even in the Markovian approximation the dissipative term is non-local in space.

The dissipation kernel, in both the non-Markovian and the Markovian cases is a complex function. As is obvious from Eqs. (\ref{eq:NMDNLS}) and (\ref{eq:MDNLS}) its real part is responsible for the decoherence effects while the imaginary part changes the oscillation frequency. The real part of $\bar{\mu} (t) = \int_0^t\mu(t-t') e^{-i(\epsilon+U)t'}dt'$ is plotted in Fig. \ref{fig:0}. This zeroth order moment constitutes the leading dissipative contribution in the rhs of Eq.~(\ref{eq:NMDNLS}). In the Markovian limit this term is the only dissipative contribution. In the case of the sub-Ohmic and the Ohmic environments and for time roughly $J t \gtrsim 1$ the difference between the Markovian and non-Markovian environments becomes very small ($<5\% $) while for the super-Ohmic case this happens earlier ($J t \gtrsim 0.25$). This figure makes the physical meaning of the Markovian limit quite clear:  Choosing for the system a coarse grained time greater than (the minimum of) the values mentioned above the environment can be safely considered as Markovian. 

Before closing the section and in connection to our numerical applications we note that the numerical solution of Eq. (\ref{eq:NMDNLS}) is by no means a trivial task as the system to be solved constitutes nonlinear integro-differential equations of the Volterra type (for a general discussion on this matter see Ref. \cite{Bahuguna2009}). We applied two different methods to tackle this problem. The first one uses the Adomian decomposition method \cite{Biazar2005} while for the second method we construct an appropriate Runge-Kutta method \cite{Filiz}. Both methods yield the same results with satisfying convergence in the time scales we will be using in the remaining of the paper. The plotted results have been produced via the Runge-Kutta method due to it's faster convergence.


\section{The Two Site BH Model}
\label{sec:4}
\begin{figure}[tb]
\centering
\includegraphics[width=8.5cm, angle=0]{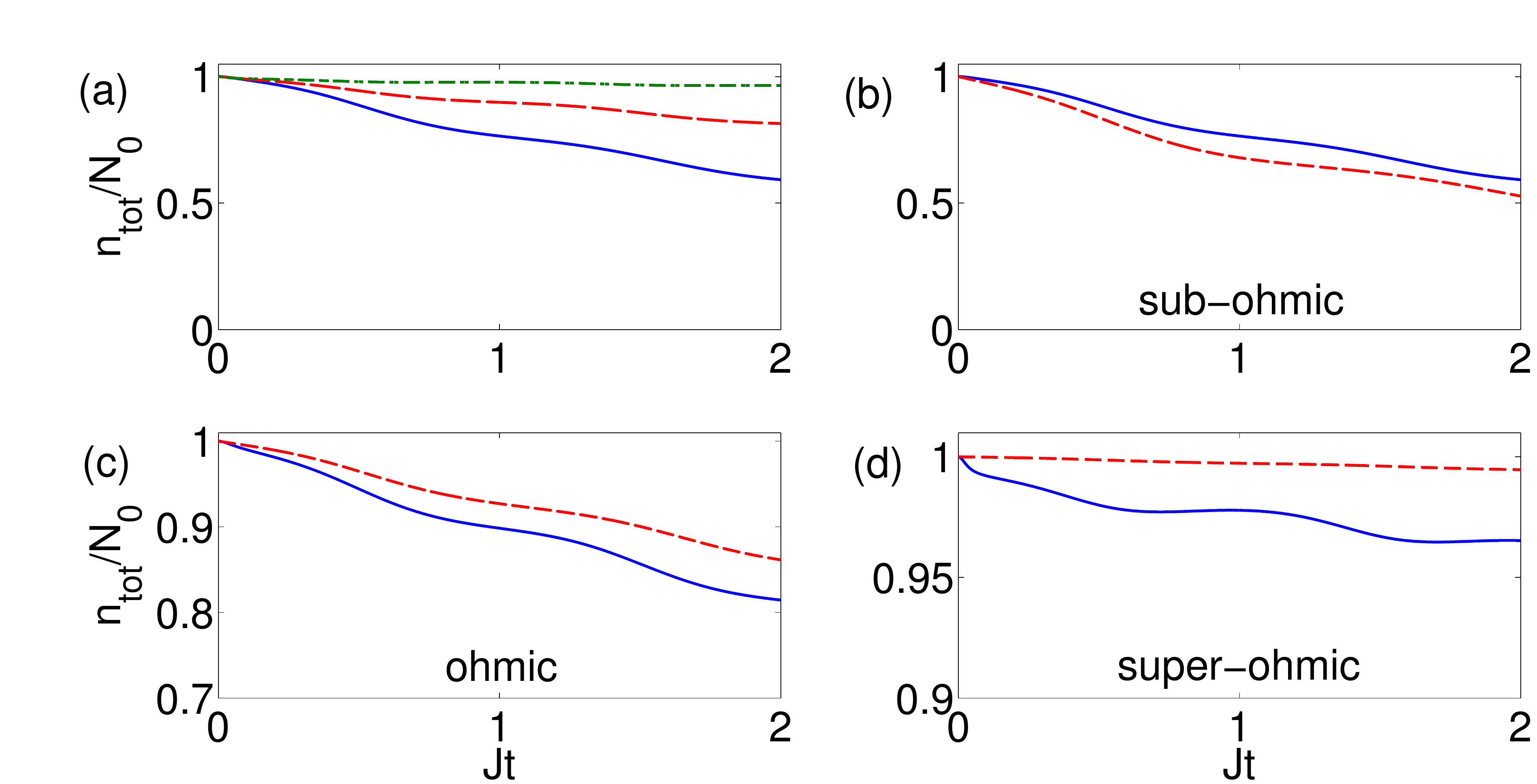}
\caption{\label{fig:01}
(Color online)
Evolution of the normalized total particle number, $n_{tot}^{norm}=n_{tot}/N_0$. (a) Comparison between sub-Ohmic, $n=0.5$ (solid blue line), Ohmic, $n=1$ (red dashed line) and super-Ohmic, $n=2$ (green dashed dotted line) non-Markovian environments. (b-d) Comparison between Markovian (red dashed line) and non-Markovian (blue solid line) dynamics for sub-Ohmic (b), Ohmic (c) and super-Ohmic (d) environments. In all cases $UN_0=6J$ and the initial condition are $(p,q)=(0.5,0.58\pi)$.
}
\end{figure}
The first example we are going to study is the two site BH model. In this case we can write the initial conditions of Eq. (\ref{eq:NMDNLS}) in the form
\begin{eqnarray}
\nonumber (a_1(0),a_2(0))&=&\sqrt{N_0}\left(\sqrt{p},\sqrt{1-p}e^{iq}\right)
\end{eqnarray}
where $p\in [0,1]$, $q\in [0,2\pi)$ and $N_0$ is the initial total particle number.

As expected, the dynamics of the system depend on the details of the environment. In Fig. \ref{fig:01} we depict the evolution of the normalized total particle number, $n_{\rm tot}^{\rm norm}=n_{\rm tot}/N_0$, for sub-Ohmic, Ohmic and super-Ohmic non-Markovian environments together with their respective Markovian limit. In all cases we have used $UN_0=6J$ and $(p,q)=(0.5,0.58\pi)$. In accordance with the behavior of the real part of the dissipation kernel (see Fig. \ref{fig:0}),  it is seen that the fastest and the slowest decays correspond to  sub-Ohmic and super-Ohmic environments, respectively.  As depicted in Fig.\ref{fig:01} (b,c,d) the differences between the Markovian and the non-Markovian cases are rather small but not negligible. For the sub-Ohmic case (Fig. \ref{fig:01}b) the difference is about $\sim 6\%$ after $Jt=2$ with the non-Markovian case having the slower decay rate. For the Ohmic, Fig. \ref{fig:01} (c), and super-Ohmic, Fig. \ref{fig:01} (d), environments the difference is about $\sim 5\%$ and $\sim 3\%$, respectively, with the Markovian case having slower decay rate. Comparing the results depicted in Fig. \ref{fig:01} (b,c,d) we see that the important factor for the decay rate is the dependence of the dissipation function on the exponent $n$. Indeed the difference between super- and sub-Ohmic cases is about $\sim 47\%$, while between sub-Ohmic and Ohmic about $\sim 33\%$. 

\begin{figure}[tb]
\centering
\includegraphics[width=8.5cm, angle=0]{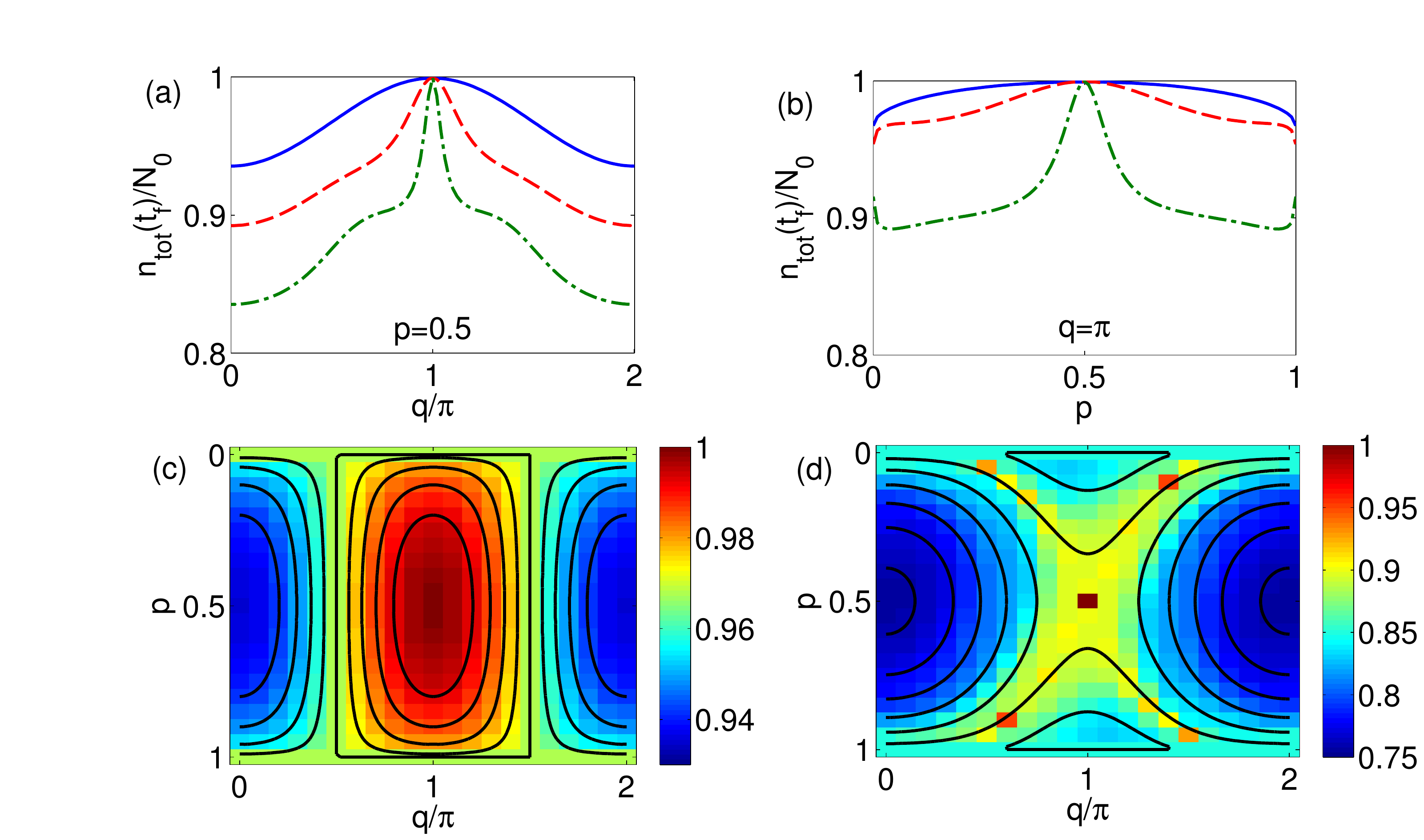}
\caption{\label{fig:02}
(Color online)
The normalized total particle number, $n_{tot}^{norm}=n_{tot}/N_0$, after a fixed propagation time $Jt_{\rm f}=2$, for different initial conditions and interaction strengths. (a,b) We have fixed $p$ and $q$ and we depict $n_{tot}^{norm}(t_f)$ as a function of $q$ and $p$ respectively, for three different interaction strengths: $UN_0=0$ (blue solid line), $UN_0=3J$ (red dashed line) and $UN_0=6J$ (green dashed dotted line). In (c) and (d) the colormap shows the value of $n_{tot}^{norm}(t_f)$ for all $(p,q)$ initial conditions, for $UN_0=0$ and $UN_0=6J$, respectively. In (c,d) the black lines depict the classical phase-space of the non-dissipative DNLS.
}
\end{figure}

\begin{figure}[tb]
\centering
\includegraphics[width=8.5cm, angle=0]{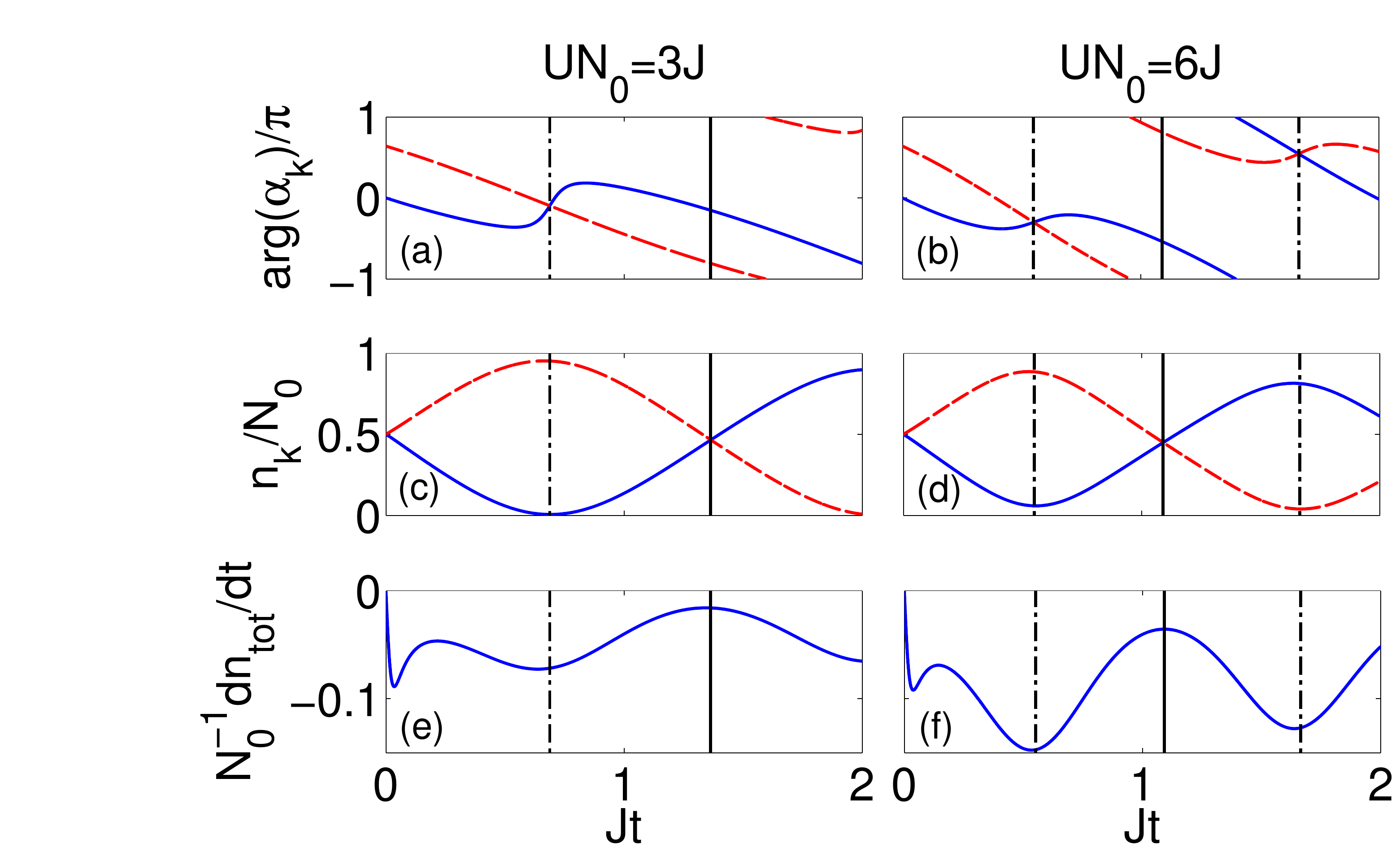}
\caption{\label{fig:03}
(Color online)
Evolution of (a,b) the phase of the fields, (c,d) the normalized fields and (e,f) decay rate, for two different interaction strengths: $UN_0=3J$ (left column) and $UN_0=6J$ (right column). The vertical black dotted dashed line depicts the time when the phase difference is zero, while the vertical black solid line the time when the population difference is zero. The initial conditions in both cases are $(p,q)=(0.5,0.64\pi)$.
}
\end{figure}

\begin{figure*}[tb]
\centering
\includegraphics[width=15cm, angle=0]{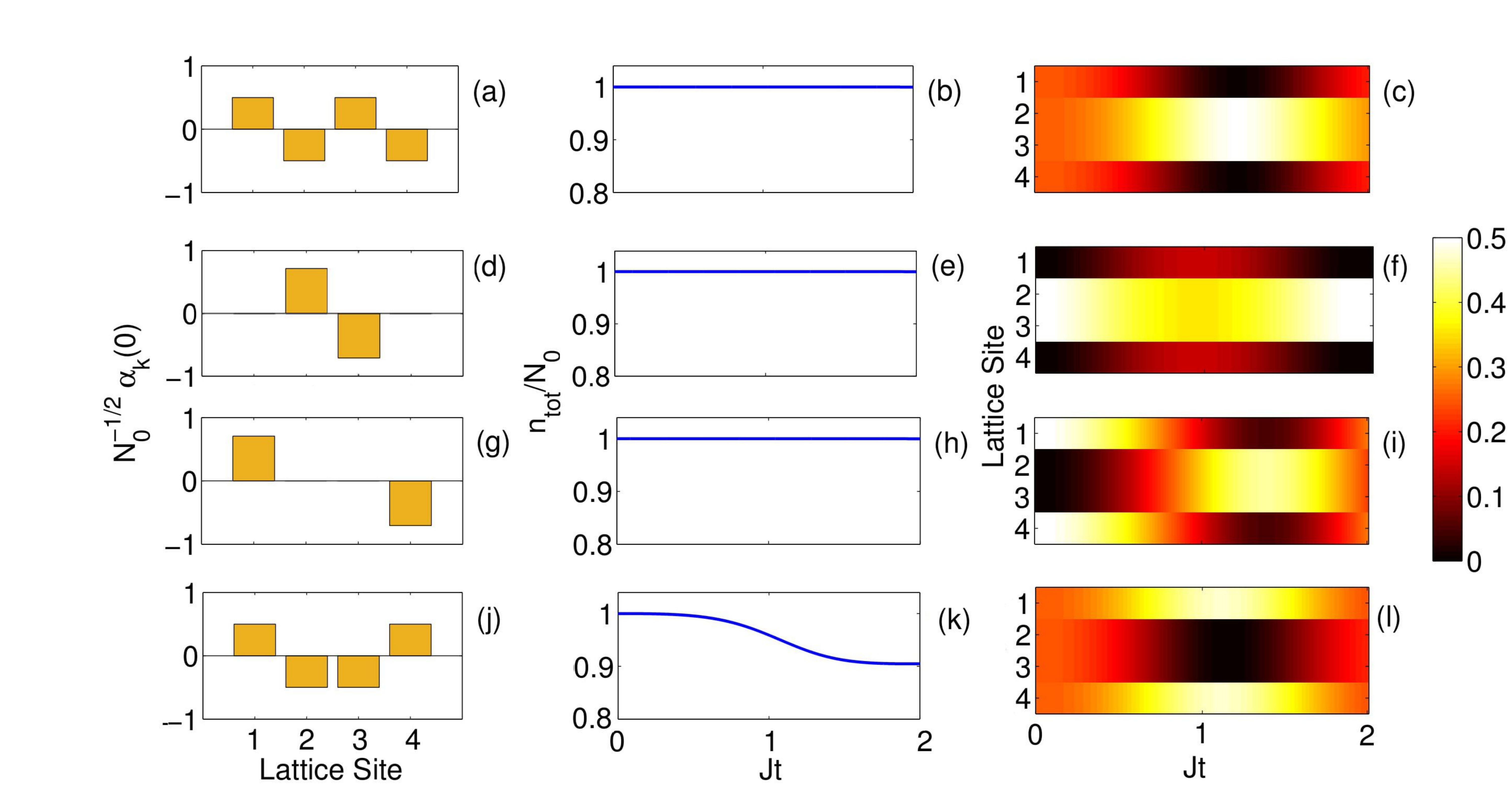}
\caption{\label{fig:04}
(Color online)
Time evolution of the normalized total particle number (second column) and of the normalized particle density in each lattice site (third column) for four different initial states (first column): (a-c) $\bm{a}(0)=\sqrt{N_0}(1,-1,1,-1)/2$, (d-f) $\bm{a}(0)=\sqrt{N_0}(0,1,-1,0)/\sqrt{2}$, (g-i) $\bm{a}(0)=\sqrt{N_0}(1,0,0,-1)/\sqrt{2}$ and (j-l) $\bm{a}(0)=\sqrt{N_0}(1,-1,-1,1)/2$. In all cases $UN_0=6J$ and we have hard wall boundary conditions.
}
\end{figure*}

An interesting question is about the way the interactions influence the decay rate. In Fig. \ref{fig:02} (a,b) we show the normalized total particle number, after a fixed propagation time, for various initial conditions and interaction strengths (providing that our approximation remains valid). Clearly, with increasing interaction strength the losses are increased. A quite interesting observation is that, in all cases, the initial condition $F_\pi=(p,q)=(0.5,\pi)$ presents zero losses. By inspecting Eq. (\ref{eq:NMDNLS}) we can confirm that $F_\pi$ is a fixed point on its lhs and, at the same time, the sum in its rhs is initially ($t=0$) zero. This observation promotes $F_\pi$ to a fixed point for the full non-Markovian DNLS and yield an evolution with zero losses. The same analysis can be carried out for the Markovian DNLS also and the zero-losses evolution appears again. In the same context, we can understand the maximum losses observed at the other fixed point of DNLS, $F_0=(p,q)=(0.5,0)$, see Fig.~\ref{fig:02} (a). Although, $F_0$ is a fixed point on the lhs of Eq. (\ref{eq:NMDNLS}), the sum on the rhs of this equation is not initially zero but attains its maximum value.
 
In Figs.~\ref{fig:02} (c) and (d) we have scanned the initial conditions and we depict $n_{\rm tot}^{\rm norm}$ after a fixed propagation time, for zero interactions and $UN_0=6J$, respectively. The results follow the classical phase space of the non-dissipative system (the black lines). In the non-interacting case, Figs.~\ref{fig:02} (c), minimum losses are observed around $F_\pi$ and maximum ones around $F_0$. In the interacting case, once again maximum losses are around $F_0$ and minimum losses along the separatrix with zero ones exactly at $F_\pi$. The above analysis is strictly related with the two site system under consideration. As we shall see in the next section the key feature of the zero loss evolution is the $Z_2$ symmetry appearing in Figs.~\ref{fig:02} (c) and (d).

To understand better why with increasing interactions we have, in general, increased losses, in Fig.~\ref{fig:03}, we have plotted the time evolution of the phases of the fields $\langle a_k\rangle$, of the population in each lattice site and the corresponding decay rate for two different interaction strengths. As we observe when the population difference is maximum, the phases of the fields are the same and we have the maximum decay rate. When the population difference is zero, the phases of the fields have the maximum difference and we have the minimum decay rate. This behavior is expected since in order to have cancellation of the dissipative term, on the rhs of Eq.~\ref{eq:NMDNLS}, at a given time instant, we must have equal populations in each site with phase difference of $\pi$. As we can see for stronger interactions we have more frequently maximum population difference between the sites, thus we have larger losses.

\section{The Four Site BH Model}
\label{sec:5}
\begin{figure*}[tb]
\centering
\includegraphics[width=15cm, angle=0]{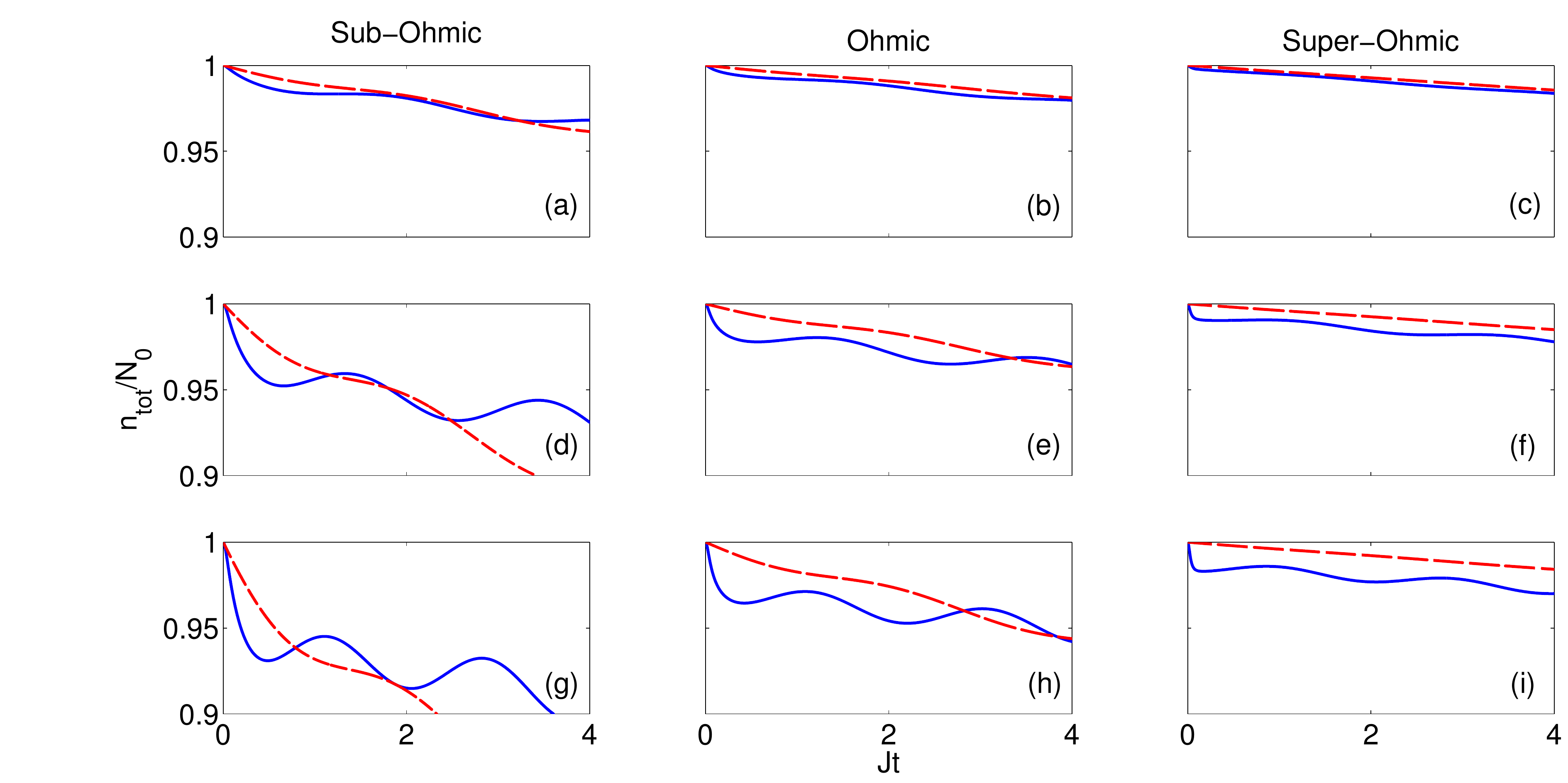}
\caption{\label{fig:05}
(Color online)
Time evolution of the normalized total particle number . In all cases the initial condition is $\bm{a}(0)=\sqrt{N_0}(1,0,0,0)$ and $UN_0=0$. For the other parameters vertically we have the Sub-Ohmic ($n = 0.5$, first column), the Ohmic case ($n = 1$, second column) and the Super-Ohmic one ($n = 2$, third column). Horizontally we have $\lambda  = 0.001$ (a-c), $\lambda  = 0.005$ (d-f) and $\lambda  = 0.01$ (g-i).
}
\end{figure*}

In the previous section we saw that the initial conditions leading to particle number conservation are fixed points of the non-dissipative DNLS. In this section we shall see that this result is exclusively  tied to the two-site case. As the forthcoming analysis will prove, the increase of  lattice sites and the corresponding enriching of the system's structure, reveals the existence of number conserving initial conditions that are not connected with fixed points. More than this,  we shall find that the increase of the number of sites enables the observation of purely non-Markovian phenomena as the return  of particles from the environment back to the system.

In Fig. \ref{fig:04} we present the four-site dynamics, for four different sets of initial conditions (first column) and we plot the evolution of the normalized total particle number $n_{\rm tot}^{\rm norm}=n_{\rm tot}/N_0$ (second column) and the normalized particle density in each lattice site (third column) for the Ohmic case and for $UN_0=6J$. In all cases the initial sum on the rhs of Eq. (\ref{eq:NMDNLS}) is zero, due to the phase difference between the initial values of $\langle a_k\rangle$. In the figures in the first, second and third row, we observe that at every time instant there are always two sites with the same population and a phase difference of $\pi$. As a consequence the total particle number is preserved. This interesting effect is connected to the $Z_2$ symmetry of the initial state. More than this, we have confirmed that every $Z_2$ invariant initial state leads to a particle-conserving evolution. 

To demonstrate the importance of the $Z_2$ symmetry we examine (see Fig. \ref{fig:04}, fourth row) an initial state in which the sum on the rhs of Eq. (\ref{eq:NMDNLS}) is zero but lacks $Z_2$ invariance. As readily confirmed the particles tunnel from the inner to the outer sites. In this process the population remains intact  but the $\pi$-phase difference disappears and particle losses appear. At the time $Jt=2$ the initial configuration is retrieved, although  with fewer particles, and the evolution continues with a step wise particle-loss profile. The effect described above, remains the same at the Markovian limit. The reason is that the crucial factor is connected with the $Z_2$ symmetry that both, the effective action and the initial state, share. For this reason,  Eqs. (\ref{eq:NMDNLS}) and (\ref{eq:MDNLS}) are $Z_2$ invariant and the same happens for the boundary conditions accompanying them. Thus, one expects the same symmetry to characterize their solution. In such a case the summation on the rhs, in both equations, yields a zero result and the particle loss disappears. 

The next quite interesting effect is a purely non-Markovian phenomenon. In Fig. \ref{fig:05} we depict the time evolution of the normalized total particle number beginning from a state in which all the particles are centered at the first site. The results present the exact dynamics as we have adopted the simplifying assumption $U=0$. The Markovian limit is characterized by a constant drop of the population as expected for a system coupled with a stochastic vacuum. In the non-Markovian cases local recoveries of the population are observed, a phenomenon that becomes increasingly important as the system-environment coupling increases. This behavior is a result of memory effects due to the non-Markovian nature of the environment. The imaginary part of the dissipation function becomes more important in the non-Markovian case and gives rise to the observed oscillations of the particle number while the real part, as already stated, gives the general losses in the population. As a consequence the particle number decays slower in the non-Markovian than in the Markovian case.

\section{Conclusions}
\label{sec:concl}

In this paper we studied the influence of a non-Markovian environment in the mean-field dynamics of a BH chain using coherent-state path integrals. Starting with a BH chain coupled to a non-Markovian vacuum, we used coherent-state path integrals to write the generating functional for the total system. Integrating out the degrees of freedom of the environment and minimizing the effective action we derived the mean-field equations for $\langle a_k\rangle$. 

With this tool at hand, we studied the dynamics of the two-site BH model. We compared the non-Markovian and Markovian dynamics for sub-Ohmic, Ohmic and super-Ohmic environments and we also compared the dynamics for different inter-particle interaction strengths between the particles. We saw that the particle loss increases as the interaction strength increases. We also investigated a quite interesting phenomenon: the initial conditions that support a time evolution without particle losses. Based on the simplicity of the two site model, we connected this particle-preserving behavior with the fixed point of the corresponding DNLS.

We have also investigated a four-site system in which the zero-loss evolution cannot be connected with fixed points. Our analysis revealed the fact that under particle  number preservation lies the fact that the effective action and the initial state of the system remain invariant under $Z_2$ rotations. Finally, we studied a purely non-Markovian effect, the return of particles from the vacuum environment back to the system, due to the non-Markovian memory effects.

The present work opens new possibilities for the study of the dynamics of many-body systems coupled to Markovian or non-Markovian environments. One of the interesting questions to be faced, is the robustness of the above presented effects against quantum fluctuations. The path integral formalism we presented makes possible the systematic examination of the quantum corrections in the effective action even in the case of strong interactions \cite{Kleinert13}. In the same framework the role of the finite temperature can be investigated via the Feynman-Vernon approach. A last but not least remark is that the above presented analysis can be extended to spin or fermionic systems through the corresponding coherent state path integrals \cite{Kor16}.


\bibliography{kordas_etal_2018}

\end{document}